\documentclass[aps,pra,twocolumn,showpacs]{revtex4}

\usepackage{graphicx}

\begin{document}

\title{Perturbative analysis of coherent quantum ratchets
in cold atom systems}

\author{M.~Heimsoth, C.E.~Creffield, and F.~Sols}

\affiliation{Departamento de F\'{\i}sica de Materiales, Facultad de Ciencias F\'{\i}%
sicas, Universidad Complutense de Madrid, E-28040 Madrid, Spain}

\date{\today}

\begin{abstract}
We present a perturbative study of the response of cold atoms in an optical
lattice to a weak time- and space-asymmetric periodic driving signal. In the noninteracting limit, and for a finite set of resonant frequencies, we show how a coherent,
long lasting ratchet current results from the interference between first and
second order processes. In those cases, a suitable
three-level model can account for the entire dynamics, yielding surprisingly
good agreement with numerically exact results for weak and moderately strong
driving.
\end{abstract}

\pacs{03.75.Kk, 05.60.Gg, 67.85.Hj}

\maketitle

\section{Introduction}

\label{intro} In recent years considerable attention has been directed at
producing directed transport in driven systems, in which the driving does
not have a net bias. This type of \textquotedblleft ratchet
effect\textquotedblright\ \cite{reimann}
is important from the viewpoint of technology in
controlling the passage of particles, ranging from electrons to nanospheres,
and also to more unusual applications such as understanding the operation of
biological molecular motors.

The most well-studied form of ratchet physics arises from the interplay
between dissipation and the driving potential. The paradigmatic example is
given by a Brownian particle in a periodic potential \cite{hanggi_review}.
The system is driven from equilibrium by periodically
varying the potential, and a ratchet current is produced when the relevant
space and time symmetries of the driven system, which would otherwise forbid
the formation of a directed current \cite{symm}, are broken.

Perhaps surprisingly, however, dissipation is not a necessary requirement for the production of a ratchet, and even in strictly Hamiltonian systems a
ratchet current can be produced
\cite{ketzmerick,tania,gong,resonance,interact,poletti,prl}. Cold atoms
loaded into optical lattices have emerged as excellent candidates to study
such coherent ratchet effects, as the level of dissipation
can be controlled rather precisely \cite{renzoni}, and indeed
such systems can be arranged to be essentially dissipation-free.
Initial investigations of ratchet-like behavior concentrated
on the \textquotedblleft quantum kicked rotor\textquotedblright , which can
be realized in experiment by subjecting a gas of ultracold atoms to a pulsed
optical lattice potential. This has allowed the detailed experimental
investigation of quantum chaos effects such as dynamical localization \cite{tania}
and quantum resonances \cite{resonance}
which can be harnessed to produce quantum coherent ratchets.

In this work we consider cold atoms subjected to an optical lattice
potential that varies smoothly with time, instead of being pulsed. We can
expect that a driving of this kind produces less heating than a kick-type
potential, and accordingly will preserve the atomic coherence better. We
consider a driving potential in which spatial and temporal symmetries can be
separately controlled, giving extreme flexibility for probing and
manipulating the system's properties. We firstly show that starting from an
unbiased initial state, symmetric in both space and time,
we are able to induce a directed current. Through a
perturbative study we find that this ratchet current arises from quantum
interference between processes which are first- and second-order in the
driving strength. We then provide a simple three-level model to describe its
properties and find excellent agreement with the exact numerical
simulations. We go on to show that the driving frequencies at which this
current occurs obeys various resonance conditions and discuss the main
features of these resonances.

\section{Asymmetric driving}

We consider a gas of cold bosonic particles held in a toroidal
trap. If the lateral dimensions of the torus are much smaller than its
radius $R$ the system becomes effectively one-dimensional. Its low
temperature dynamics are then well-described by the one-dimensional
Gross-Pitaevskii equation
\begin{equation}
H(t)=-\frac{1}{2}\frac{\partial ^{2}}{\partial x^{2}}+g\left\vert \psi
(x,t)\right\vert ^{2}+V(x,t) ~, \label{gpe}
\end{equation}%
where the short-range interaction between the atoms is described by a
mean-field term with strength $g$. The atoms are driven by a time-periodic
external potential $V(x,t)$ with zero mean, produced by modulating the
intensity of the optical lattice. We note that we measure all energies in
units of the rotational constant $\hbar ^{2}/mR^{2}$, and set $\hbar =1$.

The high level of control available in cold atom experiments allows us to
take the unusual choice of factoring the driving potential into separate
space and time components
\begin{equation}
V(x,t)=KV(x)f(t)~,  \label{Vxt}
\end{equation}
where $V(x)$ gives the spatial dependence of the optical lattice potential,
and $f(t)$ describes the time-dependence of its intensity. The archetypal
form of a symmetry-breaking ratchet potential \cite{hanggi_bartussek} is $%
V(x)=\sin (x)+\alpha \sin (2x+\phi )$, where spatial inversion symmetry is unbroken
for $\phi =\pi /2$, and is maximally broken for $\phi =0,\pi $. Accordingly
we choose to take
\begin{eqnarray}
V(x) &=&\sin (x)+\alpha \sin (2x)~, \nonumber \\
f(t) &=&\sin (\omega t)+\beta \sin (2\omega t)~,  \label{drive}
\end{eqnarray}%
where the parameters $\alpha $ and $\beta $ separately control the spatial
and temporal symmetries of the driving.
An experimental realization of such a driving potential in a cold atom system
was recently described in Ref. \onlinecite{bonn}.
The strength of the driving is
denoted by $K$, and we shall, in this paper, restrict ourselves to using
small values of $K$ for which the system shows a regular response. As $K$ is
increased the dynamics shows a rich quasiperiodic behavior, and we refer
the reader to Ref. \onlinecite{prl} for a discussion of this and its consequences.


In Ref. \onlinecite{prl} we studied the dynamics under the driving (\ref{Vxt})-(\ref{drive})
when the system
was initially prepared in the spatially-uniform, time-symmetric state
\begin{equation}
\psi (x,0)=
(2\pi)^{-1/2}~.
\label{initial-wf}
\end{equation}
This is convenient for experiment as it is the
ground state of the undriven Hamiltonian, and so can be prepared using
standard cooling techniques. Clearly the symmetry of this state prevents it
from carrying a current. We numerically integrate the wavefunction in time
using a split-operator method, in each case checking that the
time-discretization, $\Delta t$, is sufficiently small to produce converged
results. The size of $\Delta t$ depended strongly on the amplitude of
the driving, $K$, with the surprising result that smaller values of $K$
demand a much finer discretization to produce converged results.
As an additional verification, results were also checked using
a fourth-order Runge-Kutta technique.
To probe the behavior of the system we evaluate
\begin{equation}
I(t)\equiv \langle \psi
(t)|p_x|\psi (t)\rangle
\label{I-t}
\end{equation}
as a measure of the current flowing in the
ring. We note that the expectation value of the momentum operator determines
the velocity of free propagation in a time-of-flight experiment after the
optical confinement is released.

Reference \onlinecite{prl} focused on the case of driving frequency $\omega =1$.
It was found
that for small $K$ and small or moderate $g$ the current exhibits large
sinusoidal oscillations, with a period of $\sim 120$ driving periods, which
clearly averaged to a non-zero value. As the non-linearity, $g$, is increased
from zero, the oscillations in current are initially enhanced, together with
a deformation of the waveform, and then become abruptly suppressed above a
critical interaction strength. This smooth time-periodic behavior of the
current for zero or small interaction strength, clearly implies that the
driving induces an oscillation between the initial state and a single
excited state. Examining the time evolution in detail, it was found that
this oscillation occurs chiefly between $|0\rangle $ (the initial state) and
$|2\rangle $, where $|l\rangle $ denotes an eigenstate of the undriven
Hamiltonian with quantized angular momentum $l\hbar $. This suggests that,
for weak driving and weak interactions, we can model the dynamics very efficiently
by using an effective Hamiltonian operating in a reduced Hilbert space of a
small number of states. The main goal of this paper is to study the weak-driving limit in greater detail. The resulting perturbative analysis gives an excellent account of the system's ratchet dynamics.

\section{Floquet states: perturbative study.}

We focus on the behavior of a system governed by the the time-dependent
Sch\"{o}dinger equation
\begin{equation}
i\frac{\partial \psi }{\partial t}=H(t)\psi ,  \label{schrodinger}
\end{equation}
and we will concentrate on the non-interacting case, $g=0$, with an initial wave function (\ref{initial-wf}).
The time-periodicity of the Hamiltonian $H(t)=H(t+T)$ implies that solutions
of the Schr\"{o}dinger equation are given by time-periodic functions known
as Floquet states, analogous to the Bloch wave solutions familiar from
studies of spatially-periodic systems. Floquet states are of the general form%
\begin{equation}
\psi_j(x,t)=\exp (-i\varepsilon_jt)\phi_j(x,t)~,  \label{Floquet}
\end{equation}
where $\phi_j(x,t)=\phi_j(x,t+T)$ is a time-periodic function and the
quasienergy $\varepsilon _{j}$ is defined modulo $2\pi /T\equiv \omega$.

We also note that our Hamiltonian is of the form%
\begin{equation}
H(t)=H_{0}+V(t),
\end{equation}%
where $V(t)$ is given by the driving (\ref{Vxt})-(\ref{drive}). For small $K$
we expect that a suitable perturbative study will yield the Floquet states
from the eigenstates of the unperturbed Hamiltonian, which satisfy%
\begin{equation}
H_{0}|l\rangle =\epsilon _{l}|l\rangle ~,
\end{equation}
where $|l\rangle $ is an eigenstate of the momentum along the ring with
space-periodic boundary conditions, with the wavefunction
\begin{equation}
\langle x|l\rangle =(2\pi)^{-1/2}\exp (ilx)
\ ,
\end{equation}
and unperturbed energy $\epsilon _{l}=l^{2}/2$.

We note first that the time-periodic functions satisfy the equation
\begin{equation}
\lbrack -i\partial _{t}+H(t)]\phi_j(t)=\varepsilon _{j}\phi_j(t)~.  \label{H-u}
\end{equation}%
Since we are interested in the case of space-periodic boundary conditions, $%
\phi_{j}(x+2\pi ,t)=\phi_{j}(x,t)$ and $V(x+2\pi ,t)=V(x,t)$, we may exploit the
combined periodicity in $x$ and $t$ (i.e. on the square $[0,2\pi ]\times
\lbrack 0,T]$) and map the dynamics into a time-independent problem in an
effective two-dimensional geometry. Specifically, we make the replacement $%
t\rightarrow y$ \ and rewrite Eq. (\ref{schrodinger}) as an eigenvalue
equation
\begin{equation}
\left[ p_{x}^{2}/2m+p_{y}+V(x,y)\right] \phi _{j}(x,y)=\varepsilon _{j}\phi
_{j}(x,y)  \label{eqn: making time spatial}
\end{equation}
where $p_{y}=-i\partial /\partial y$. The generalized Hamiltonian defined in (\ref{eqn:
making time spatial}) is Hermitian. Its only anomaly is that it is
not bounded from below, but this does not prevent us from applying standard
tools that do not rely on the existence of a minimum energy. So finding the
Floquet states (\ref{Floquet}) satisfying (\ref{schrodinger}) amounts
to solving the Hermitian eigenvalue problem (\ref{eqn: making time spatial}).

We recall that the driving,
\begin{equation}
V(x,y)=K[\sin (x)+\alpha \sin (2x)][\sin (\omega y)+\beta \cos (2\omega y)]~,
\label{Vxy}
\end{equation}
is small, so a useful starting point is given by the unperturbed Floquet
states in the $xy$ representation, which satisfy%
\begin{equation}
(p_{x}^{2}/2m+p_{y})\phi _{lm}^{0}(x,y)=\varepsilon _{lm}^{0}\phi
_{lm}^{0}(x,y)~,  \label{H-not}
\end{equation}
where%
\begin{eqnarray}
\phi _{lm}^{0}(\vec{r})&=&(2\pi T)^{-1/2}\exp (i\vec{k}_{lm}\cdot \vec{r}%
) \nonumber \\
&=&(2\pi T)^{-1/2}\exp (ilx-i\omega _{m}y)~,  \label{plane-wave}
\end{eqnarray}
with $\vec{r}=(x,y),$ and  $\vec{k}_{lm}=(l,-\omega _{m})$,
where $\omega _{m}=$ $2\pi
m/T=m\omega $, and $l,m$ are integers. The
zeroth-order approximation to the Floquet quasienergies
is given by
\begin{equation}
\varepsilon _{lm}^{0}=\frac{l^{2}}{2}-\omega _{m}~.
\end{equation}

The normalization has been chosen to satisfy orthonormality,
\begin{equation}
\int_{0}^{T}dy\int_{0}^{2\pi }dx~[\phi _{lm}^{0}(x,y)]^{\ast }\phi
_{l^{\prime }m^{\prime }}^{0}(x,y)=\delta _{ll^{\prime }}\delta _{mm'}~.
\end{equation}

In the following we focus on those values of $l$ and $m$ which satisfy the
\textit{resonance condition}%
\begin{equation}
\omega _{m}=\frac{l^{2}}{2}\,\,,~~\mbox{i.e.}~~\varepsilon _{lm}^{0}=0~,
\label{resonance}
\end{equation}
where we expect to find the highest values of the ratchet current.

\begin{figure*}
\includegraphics[width=0.7\textwidth,clip=true]{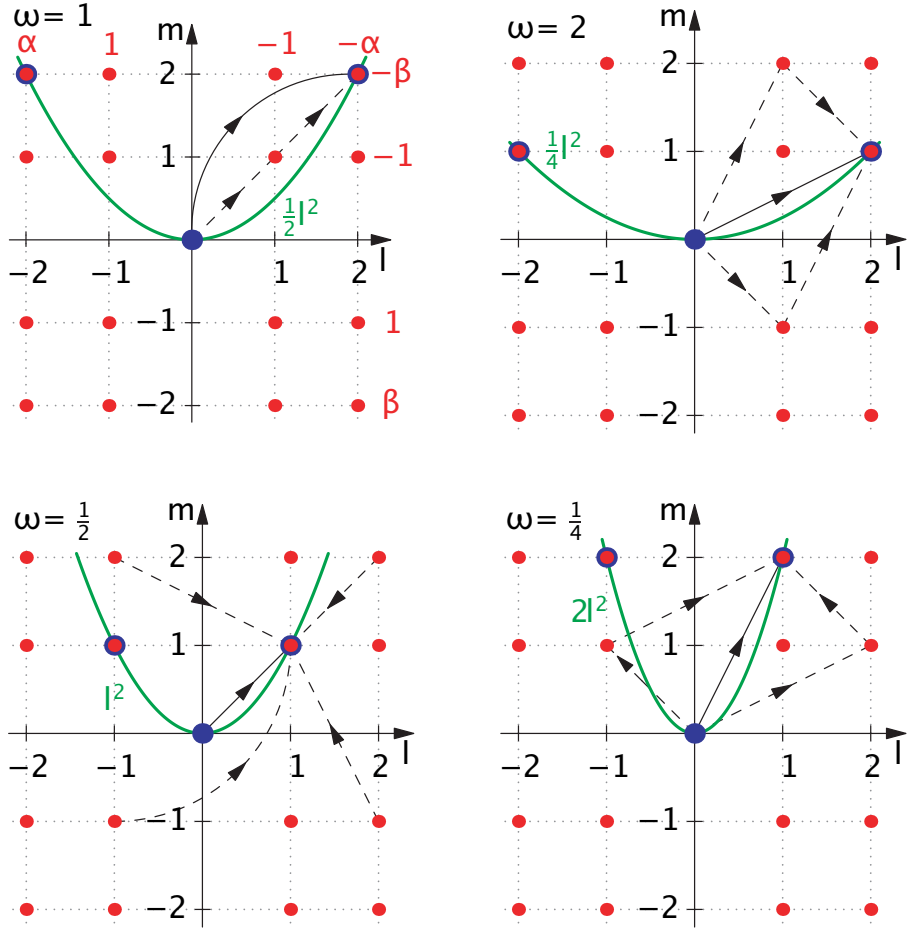}
\caption{Schematic representation of the driving-induced transitions for driving frequencies
$\omega=1/4,1/2,1,2$; see main text for the meaning of the various symbols.
In each case the direct, first-order (solid arrows) and indirect, second-order (dashed arrows) transition amplitudes interfere to produce a ratchet current. For clarity, in the $\omega=1/2$ figure the first dashed arrow of the various second-order processes has been omitted.}
\label{gspaceomega-four}
\end{figure*}

The periodic driving (\ref{Vxy}) has a finite number of Fourier components,
so that in the expansion%
\begin{equation}
V(\vec{r})=\sum_{\vec{g}}V_{\vec{g}}\exp (i\vec{g}\cdot \vec{r})
\label{V-Fourier}
\end{equation}
only a handful of reciprocal lattice vectors $\vec{g}$ satisfy $V_{\vec{g}%
}\neq 0$. Specifically, the driving Fourier component is nonzero for the 16
values of $\vec{g}$ represented by small red dots in any of the four figures
shown in Fig. \ref{gspaceomega-four}. In each of these figures, the blue
circles indicate the $\vec{k}_{lm}$ vectors that satisfy the resonance
condition (\ref{resonance}), which can also be written as%
\begin{equation}
m=l^{2}/2\omega ~,  \label{resonance-1}
\end{equation}
and whose continuous version is represented by the green parabola.

In this peculiar $xy$ space all resonant states are \textquotedblleft
degenerate\textquotedblright\ with energy $\varepsilon _{lm}^{0}=0$. These
states are connected by the driving (\ref{Vxy}), (\ref{V-Fourier}). By
construction, our initial state corresponds to the unperturbed Floquet state
$|00\rangle ,$ where in general\ $|lm\rangle $ is the state characterized by
$\vec{k}_{lm}$, its wave function being%
\begin{equation}
\langle x,y|lm\rangle =\phi _{lm}^{0}(x,y)~.
\end{equation}
This state $|00\rangle$ is represented by the filled blue circle
in each of the four graphs of
Fig. \ref{gspaceomega-four}. Thus we expect the driving to mix $|00\rangle $
with the other resonant Floquet states satisfying $\varepsilon_{lm}^{0}=0$.

Let us focus first on the case $\omega =1$ (upper-left
graph of Fig. \ref{gspaceomega-four}). Starting from the state%
\begin{equation}
\phi _{00}^{0}(x,y)=(2\pi T)^{-1/2}~,
\end{equation}
we expect the system to evolve from $|00\rangle $ towards $|22\rangle $ and $%
|\bar{2}2\rangle $, where $\bar{l}\equiv -l$. In general, under the effect
of driving the state $|00\rangle $ will also mix with higher-lying
unperturbed Floquet states, but we may expect that mixing to be weaker as it
involves higher powers in the Fourier components $V_{\vec{g}}$, all of which
are small because $K$ is assumed to be small. Thus a truncated Hilbert space
spanned only by the Floquet states $\left\{ |00\rangle ,|22\rangle ,|2\bar{2}%
\rangle \right\} $ may suffice to describe the dynamics under weak driving.
The next section is devoted to the formulation of the three-level model.


\section{Three-level model: effective matrix elements.}

Inspection of the upper-left Fig. \ref{gspaceomega-four} strongly suggests
that for $\omega =1$, the mixing between $|00\rangle $ and e.g. the state $%
|22\rangle $ must be dominated by the interference between the first-order
process involving $\vec{g}=(2,2)$ once, and the the second-order process
involving $\vec{g}=(1,1)$ twice (with $\vec{g}$ given in the units of the
array there shown), while visiting the state $|11\rangle $ virtually. In this
section we calculate the effective matrix elements which account for
this first- and second-order mixing in an effective time-independent problem
defined in the $xy$ space.

First we note that, in this $xy$ representation, we deal with a formally
time-independent problem defined by the Hamiltonian%
\begin{equation}
H=H_{0}+V~,
\end{equation}
where $H_{0}$ is given in (\ref{H-not}) and $V$ by (\ref{Vxy}). The full
Green function $G(z)\equiv (z-H)^{-1}$ can be related to its unperturbed counterpart
$G^{0}(z)\equiv (z-H_{0})^{-1}$ through the Dyson equation \cite{economou}
\begin{equation}
G=G^{0}+G^{0}VG~.
\end{equation}
One may also describe the effect of the perturbation $V$ in terms of the
$T$-matrix (not to be confused with the time-period) satisfying%
\begin{equation}
T=V+VG^{0}T
\end{equation}
or, equivalently,
\begin{equation}
G=G^{0}+G^{0}TG^{0} .
\end{equation}
Thus the effective dynamics up to second order in $V$ may be described in
terms of an effective $T$-matrix approximated as%
\begin{equation}
T(z)\simeq V+VG^{0}(z)V
~.
\label{approx-T}
\end{equation}
Finding the matrix elements of this approximate, second-order $T$-matrix is
equivalent to finding those of an effective $\tilde{V}$ which, treated to
first order, yields the correct second-order dynamics. So the next goal is
to compute the matrix elements $\langle lm|T|l^{\prime }m^{\prime }\rangle .$
Applying the closure relation to (\ref{approx-T}) we obtain%
\begin{equation}
\langle j|T(z)|j^{\prime }\rangle = \langle j|V|j^{\prime }\rangle
+\sum_{j^{\prime \prime }}\frac{\langle j|V|j^{\prime \prime }\rangle
\langle j^{\prime \prime }|V|j^{\prime }\rangle }{z-\varepsilon _{j^{\prime
\prime }}^{0}}~,  \label{matrix-element}
\end{equation}
where $j$ is a short-hand notation for the quantum numbers $lm$.

For the case $\omega =1$ we restrict our analysis to the matrix elements
between the
three resonant states $\{|00\rangle ,|22\rangle ,|\bar{2}2\rangle \}$.
These three states have an unperturbed energy $\varepsilon _{j}^{0}=0$, so
in (\ref{matrix-element}) we focus on the shell $z=0$. Inspection of Fig.
\ref{gspaceomega-four} clearly shows that the intermediate state $|j^{\prime
\prime }\rangle $ connecting $|00\rangle $ and $|22\rangle $ is $|11\rangle $%
, while $|\bar{1}1\rangle $ is the intermediary between $|00\rangle $ and $|%
\bar{2}2\rangle $. Thus, for example,
\begin{eqnarray}
\langle 00|T|22\rangle &=& \langle 00|V|22\rangle -\frac{\langle 00|V|11\rangle
\langle 11|V|22\rangle }{\varepsilon _{11}^{0}}
=V_{22}-\frac{V_{11}^{2}}{%
\varepsilon _{11}^{0}} \nonumber \\
&=& \frac{K}{4}\left( -\alpha \beta +\frac{K}{2}\right)
\equiv \Gamma _{+}~,
\end{eqnarray}%
where $V_{lm}$ stands for the Fourier component $V_{\vec{g}}$, with $\vec{g}=(l,m)$ given in the units of Fig. \ref{gspaceomega-four}.

Similarly,
\begin{equation}
\langle 00|T|\bar{2}2\rangle =\Gamma _{-}~,
\end{equation}
with%
\begin{equation}
\Gamma_{\pm }=\frac{K}{4} \left( \mp \alpha \beta +\frac{K}{2} \right) ~.
\end{equation}

Restricting to these two well-behaved second-order matrix elements, we
construct a three-level model defined by the Hamiltonian matrix
\begin{equation}
H_{3}\equiv \left[
\begin{array}{ccc}
0 & \Gamma _{+} & 0 \\
\Gamma _{+} & 0 & \Gamma _{-} \\
0 & \Gamma _{-} & 0%
\end{array}%
\right] ~,  \label{H-3}
\end{equation}
spanning the space $\left\{ |22\rangle ,|00\rangle ,|2\bar{2}
\rangle \right\}$ or, for brevity,
$\left\{ |2\rangle ,|0\rangle ,|\bar{2} \rangle \right\}$
(using this ordering). In the subspace $\left\{ |2\rangle ,|\bar{2}\rangle
\right\} $ it is always possible to introduce a rotation such that one state
is decoupled from $|0\rangle $. Specifically, if we define the orthonormal states
\begin{eqnarray}
|a\rangle &=&\frac{1}{\sqrt{\Gamma _{+}^{2}+\Gamma _{-}^{2}}}\left( \Gamma
_{+}|2\rangle +\Gamma _{-}|\bar{2}\rangle \right) \label{state-a} \\
|b\rangle &=&\frac{1}{\sqrt{\Gamma _{+}^{2}+\Gamma _{-}^{2}}}\left( \Gamma
_{-}|2\rangle -\Gamma _{+}|\bar{2}\rangle \right) ~, \label{state-b}
\end{eqnarray}%
we find that $\langle 0|T|b\rangle =0,$ so that (\ref{H-3}) transforms into%
\begin{equation}
\tilde{H}_{3}=\left[
\begin{array}{ccc}
0 & \Gamma & 0 \\
\Gamma & 0 & 0 \\
0 & 0 & 0%
\end{array}%
\right] ~,  \label{H3-tilde}
\end{equation}
where%
\begin{equation}
\Gamma =\langle 0|T|a\rangle =\sqrt{\Gamma_{+}^{2}+\Gamma _{-}^{2}}
=\frac{K}{4\sqrt{2}}\sqrt{K^2+4\alpha^2\beta^2}
~.
\end{equation}
Therefore, for each particular driving the general three-level model can be truncated to an effective two-level problem, as found numerically in Ref. \onlinecite{prl}.

\section{Averages of the current}

Returning to the real time ($xt$) picture, it is clear that if the system is
initially prepared in a given state $\psi (x,0)$, its subsequent evolution
for $t>0$, when the driving is on [we assume (\ref{Vxt}) is multiplied by a
step function $\theta (t)$], can be economically written as an expansion in
the complete basis of the Floquet states, of general form (\ref{Floquet})
\begin{eqnarray}
\psi (x,t) &=& \sum_{j}c_{j}\exp (-i\varepsilon _{j}t)\phi _{j}(x,t)~,
\label{Floquet-expansion}\\
c_{j} &=& \int \phi _{j}(x,0)^{\ast }\psi (x,0)dx~,
\label{eqn: expansion coefficients}
\end{eqnarray}
In this sense the Floquet states represent a generalization
of the standard energy eigenstates of static Hamiltonians
to the case of time-periodic systems.

Within our perturbative approach, we have
seen that the three-level picture reduces in each case to a two-state ($%
|0\rangle $ and $|a\rangle $) problem, of Hamiltonian (\ref{H3-tilde}).
Thus, if the initial state is $|\psi (0)\rangle =|0\rangle $, or $\psi
(x,0)=(2\pi T)^{-1/2}$, the system will undergo a simple oscillation
of the Rabi type between
the unperturbed Floquet states $|0\rangle $ and $|a\rangle $.
The frequency of those oscillations will be $2\Gamma $.
Oscillations of exactly this type were observed in the numerical
investigation of this model in Ref. \onlinecite{prl} and
in the recent experimental investigation \cite{bonn}.

As well as evaluating the time-dependent current, it is also convenient to
calculate the {\em time-averaged} current. This is particularly
important to verify the existence of a long-lasting ratchet effect, which requires that the time-averaged current
remains non-zero as the observation period tends to infinity.
Such time-averages can be formed in two distinct ways.
The first is the {\em stroboscopic} average, in which the current is
evaluated only at discrete times $t_n = t_0+nT$, where $t_0 \in \left( 0,T \right]$. It can be defined as
\begin{equation}
\bar{I}_s(t_0,N)  \equiv
\frac{1}{N}\sum_{n=0}^{N}I(t_0+nT)~.
\label{strobo-average}
\end{equation}
This is frequently the
scheme of measurement most convenient for experiment.
One example is when
the driving potential $V(x,t)$ is obtained by periodically
accelerating and decelerating the optical lattice, thereby producing an
inertial force in the rest frame of the lattice. Measurements of the
current, however, are made in the rest frame of the laboratory,
and so it is convenient to make measurements at times when
the laboratory and lattice rest frames coincide, which
occurs stroboscopically. The other averaging
scheme is a {\em continuous} time average,
\begin{equation}
\bar{I}_c (\tau)\equiv  \tau^{-1} \int_0^{\tau} I(t) dt ~.
\label{cont-average}
\end{equation}
When the period of the driving is much shorter than the response
of the system, these averages will coincide at long times.
However, for lower driving frequencies
it is possible that the two time-averages will yield different results,
as the stroboscopic sampling will only capture a subset of the system's
dynamics.

At long times ($N,\tau \rightarrow \infty$), we can use (\ref{I-t}), (\ref{Floquet-expansion}), (\ref{strobo-average}), and (\ref{cont-average}) to write these time-averaged currents as
\begin{eqnarray}
\bar{I}_s(t_0) &=& \sum_{j}|c_{j}|^{2}\langle \phi
_{j}(t_0)|p_x|\phi _{j}(t_0)\rangle ~,
\label{eqn: time averaged currents} \\
\bar{I}_c &=&
\frac{1}{T}\int_{0}^{T}dt_0 \, \bar{I}_s(t_0)\;.
\label{eqn_continuous}
\end{eqnarray}
In (\ref{eqn: time averaged currents}) we have assumed that $\varepsilon_j \neq \varepsilon_{j'}$ for all $j\neq j'$.

Once the driving has been switched on and Rabi
oscillations have started, the system will spend, on average, half its time
in $|0\rangle $ and $|a\rangle $. Thus the analytical prediction for the
ratchet current is simply ($I\equiv \bar{I}_c$)
\begin{equation}
I=I_{a}/2 ~,
\end{equation}
where \thinspace $I_{a}\equiv \langle a|p_{x}|a\rangle $ is the current
carried by $|a\rangle $.

We first consider the case of the main resonance $\omega=1$.
In this case the Rabi oscillation occurs between $|0\rangle$ and
$|a\rangle$ given by (\ref{state-a}), with frequency
\begin{equation}
\Omega_R=2\Gamma=\frac{K}{2\sqrt{2}}\sqrt{K^2+4\alpha^2\beta^2}~.
\label{Rabi-frequency}
\end{equation}
We note that, in our reduced units, the states $|2\rangle$
and $|\bar{2}\rangle$, carry currents 2 and -2, respectively. As a result, the
ratchet current can be shown to be
\begin{equation}
I=\frac{4K\alpha \beta }{K^{2}+4\alpha ^{2}\beta ^{2}}~.
\label{ratchet-current}
\end{equation}%
From (\ref{ratchet-current}) the beautiful picture emerges of a coherent
ratchet current stemming from the interference between first- and
second-order processes creating an imbalance between the matrix elements
coupling the initial, time-symmetric state $|0\rangle $ to the current-carrying states $|2\rangle $ and $|\bar{2}\rangle $. Most
importantly, the ratchet current exists only if both $\alpha $ and $\beta $
are nonzero, i.e. if the driving is \textit{both} space- and
time-asymmetric, in agreement with the symmetry analysis provided in
\cite{denisov}.
We can further observe that the ratchet
current is a unique function of the product $\alpha \beta $,
and for this driving frequency the ratchet current is maximized
for $\alpha \beta = K/2$, in excellent agreement with the experimental observation \cite{bonn}.

\begin{center}
\begin{figure}
\includegraphics[width=0.40\textwidth,clip=true]{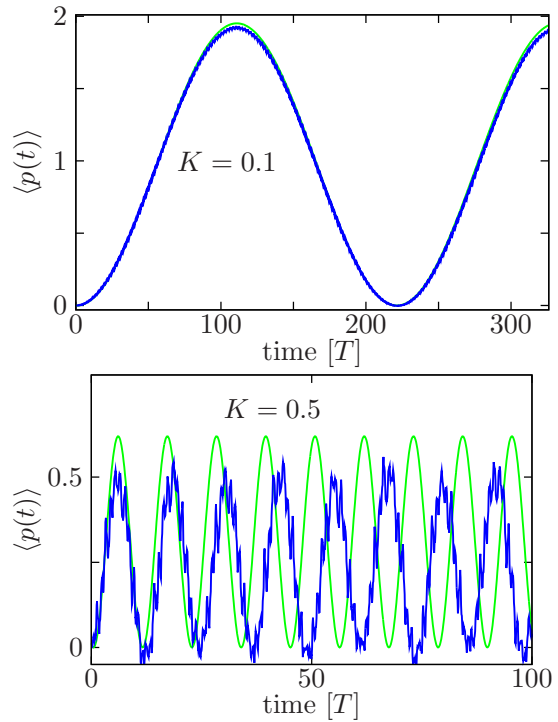}
\caption{Comparison of the time-dependent current predicted by
the effective three-level model and the exact numerical
results, for $\omega=1$ and asymmetry parameters $\alpha = \beta = 0.2$.
The three-level model predicts Rabi oscillations which fit the
exact results extremely well for weak driving ($K=0.1$). For strong
driving ($K=0.5$) the exact results show additional small, high-frequency oscillations,
but the main behavior is still reasonably well-described by the effective model.}
\label{rabi}
\end{figure}
\end{center}

Figure \ref{rabi} shows the extremely good agreement between the
analytical prediction of the three-level model and the full numerical
simulation for the time-dependent ratchet current in the weak coupling case (%
$K=0.1$). In particular, this means that the analytical predictions (\ref{Rabi-frequency}) and (\ref{ratchet-current}) for the Rabi frequency and the long-time ratchet current become essentially exact in the weak-driving limit. As expected, the agreement with the perturbative analytical
calculation worsens for higher $K$.

\begin{center}
\begin{figure}
\includegraphics[width=0.40\textwidth,clip=true]{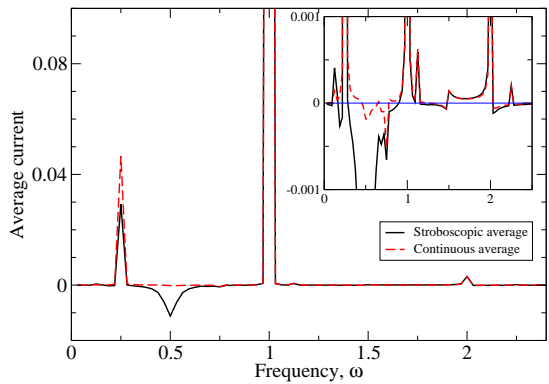}
\caption{Time-averaged current, averaged over 4000 driving periods
for a weakly-driven system ($K=0.05$, asymmetry parameters $\alpha=\beta=0.2$), plotted as a function of the driving frequency.
Four peaks appear, at driving frequencies of $\omega=0.25, 0.5,1,2$
in agreement with the resonant condition (\ref{resonance}).
{\em Inset}: Magnified view, showing the existence of sub-resonances at
commensurate fractions of the resonant frequencies. We also note
the reduction of the $\omega=0.5$ resonance when the time-average
is evaluated continuously instead of stroboscopically.}
\label{scan}
\end{figure}
\end{center}

We recall that our analysis predicts, in addition to the main resonance
at $\omega=1$, additional resonances for $\omega=1/4, 1/2, \mbox{ \ and \ }2$.
To verify this prediction we show in Fig. \ref{scan}
the time-averaged current obtained for a fixed driving strength of $K=0.05$
as the frequency is varied over a wide range.
We can first note that we indeed see peaks at the four resonant frequencies
indicated by our model.
A similar set of resonance peaks was observed in the
experimental investigation of this ratchet \cite{bonn}.
The $\omega=1$ peak is considerably larger
than the others, and it is interesting to note that the response
for $\omega=1/2$ is of opposite sign to the other peaks.
This peak is also unusual in its sensitivity to the
averaging procedure used; the stroboscopic result
is larger and broader than the continuous time-average.
In the inset we show an enlargement of the small-scale structure
in the current response, which indicates the existence
of further families of sub-resonant peaks, of much smaller
magnitude than the four primary peaks, occurring at commensurate
fractions of the driving frequencies. This perhaps indicates
the role of higher order interference processes, which could in principle
be described by a suitable generalization of our procedure.

\begin{center}
\begin{figure}
\includegraphics[width=0.40\textwidth,clip=true]{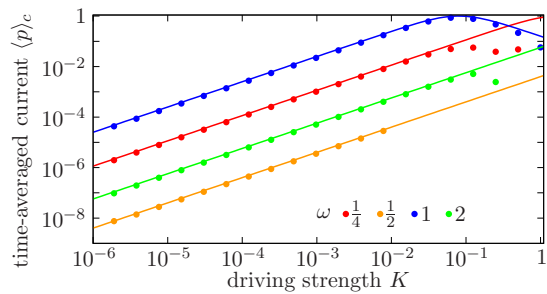}
\caption{Comparison of the analytical results with numerically-exact data
for the four principal resonances, $\omega=0.25,0.5,1,2$ 8 (see Table I). The values
for $\omega=\frac{1}{2}$ are actually negative,
and so for convenience we plot their absolute value.
For all the curves the asymmetry parameters were set to be
$\alpha = \beta = 0.2$.
We can see that for weak driving strengths the agreement is
excellent; the effective three-level model produces
{\em quantitatively} accurate results. For higher driving
strengths the model diverges from the exact results,
as expected for a perturbative result.}
\label{logcurrent}
\end{figure}
\end{center}

In the same way as for the $\omega=1$ resonance, we can obtain analytical
results for the ratchet currents produced by the other main resonances,
governed by first and second order
transitions as indicated in Fig. \ref{gspaceomega-four}.
These results are given in Table \ref{analytic}.
In Fig. \ref{logcurrent}
we compare the analytical and numerical predictions for the
continuously time-averaged ratchet current for the four main resonances. The
agreement is excellent. We note in particular the linear behavior
$I\propto K$ for small $%
K,$ which is also predicted analytically, and the existence of a
maximum current for $\omega=1$, occurring for a driving strength
of $K =0.08$.

\begin{center}
\begin{figure}
\includegraphics[width=0.40\textwidth,clip=true]{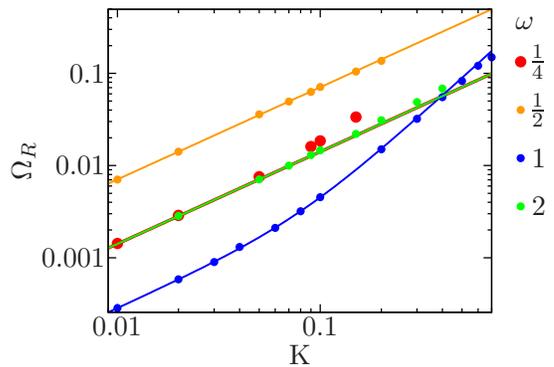}
\caption{Rabi frequencies, $\Omega_R$, predicted
by the effective three-level model [see Eq. (\ref{Rabi-frequency}) and Table \ref{analytic}]
are shown with solid lines, to compare with data extracted
from numerically exact simulations. As in Fig. \ref{logcurrent}, we see
excellent agreement between analytics and numerics.}
\label{frequencycomparison}
\end{figure}
\end{center}

As well as the amplitude of the ratchet current, the effective
model also provides predictions for the period of the Rabi oscillations.
In Fig. \ref{frequencycomparison} we compare these predictions [see Eq. (\ref{Rabi-frequency})] with the numerically
exact results, and again see excellent, quantitative
agreement over a wide range of driving amplitudes.

\begin{center}
\begin{figure}
\includegraphics[width=0.40\textwidth,clip=true]{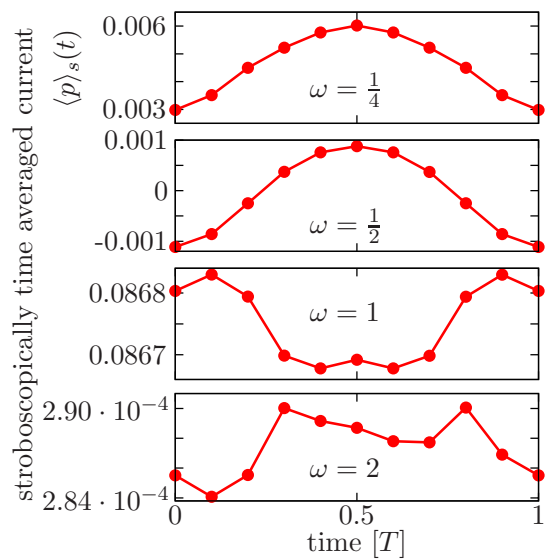}
\caption{The stroboscopic average is evaluated at discrete
times $t_n = t_0 + n T$. Here we show the dependence of
this time average on the initial time sampling point $t_0$.
For $\omega=0.25,1,2$ the variation is weak, but
for $\omega=0.5$ the dependence is strong, and contributes
to the low value of the continuous current average
for this value of the driving frequency. The number of driving periods used to compute the stroboscopic average is, from top to bottom, 400, 400, 600, 300. This amounts to about ten complete Rabi periods ($2\pi/\Omega_R$) in the first two cases, and one Rabi period in the last two.}
\label{start}
\end{figure}
\end{center}

Finally, in Fig. \ref{start} we show the stroboscopically-averaged
current
as a function of the initial sampling point time $t_0$.
For $\omega=0.25,1,\mbox{ \ and \ }2$ the stroboscopic quantity
shows only a very weak dependence on $t_0$.
The $\omega =0.5$ resonance, however, does
display an important dependence on $t_0$, and indeed changes sign
as $t_0$ varies over the range $0 \leq t_0 \leq T$.
This causes the continuously-averaged current to be significantly
smaller than a stroboscopic estimate for this driving frequency [see Eq. (\ref{eqn_continuous})],
as can be seen in Fig. \ref{scan}.

\begin{table}
\begin{centering}
\begin{tabular}[t]{ccc}
\hline
\phantom{aa}$\omega$\phantom{aa} & $\Omega_R$ & $I$ \\
\hline
$\frac{1}{4}$ & $\frac{K}{\sqrt{2}}\left[(\frac{4}{7}K\alpha)^2+\beta^2\right]^{1/2}$ & $ \frac{56K\alpha\beta}{7\beta^2+16(K\alpha)^2}$ \\
$\frac{1}{2}$ & \phantom{a}$\frac{K}{\sqrt{2}}\left[1+(\frac{1}{10}K\alpha\beta)^2\right]^{1/2}$\phantom{a} & $-\frac{10K\alpha\beta}{(K\alpha\beta)^2+100}$ \\
$1$           & $\frac{K}{2\sqrt{2}}\left[K^2+(2\alpha\beta)^2
\right]^{1/2}$        & $\frac{4K\alpha\beta}{K^2+(2\alpha\beta)^2}$ \\
$2$           & $\frac{K}{\sqrt{2}}\left[(\frac{1}{35}K\beta)^2+\alpha^2\right]^{1/2}$
& $\frac{70K\alpha\beta}{(35\alpha)^2+(K\beta)^2}$ \\\hline
\end{tabular}
\caption{Analytic results for the Rabi frequencies and the continuously time-averaged ratchet currents for the four resonant frequencies considered.}
\label{analytic}
\end{centering}
\end{table}

\section{Conclusions}
We have studied an unusual form of flashing ratchet, in which the
spatial and time symmetries can be controlled independently.
Using a novel form of perturbation theory, we find that we
are able to describe the weak-driving regime of this
system to a surprisingly high degree of accuracy by using a three-level
effective model, which later reduces, in each particular case, to a simple two-level model. This provides
an analytical underpinning to the phenomenological two-level
model introduced in Ref. \onlinecite{prl} to describe this ratchet system.
The ratchet current arises from the interference
between first- and second-order driving-induced processes, and so is a
purely quantum coherent effect, not involving dissipation.
It should be noted that we have neglected the effect of the
non-linear interaction $g$. In Ref. \onlinecite{prl} it was shown that
this has the effect of damping the Rabi oscillations, eventually producing
a ``self-trapped'' state. Introducing the non-linearity in a consistent
way to this model remains a interesting topic for future work.

\bigskip
{\em Acknowledgments}. This work was supported by MICINN (Spain),
through grant FIS-2007-65723 and the Ram\'on y Cajal Program (CEC).

\end{document}